\documentclass[usegraphicx]{mn2e}
\usepackage{txfonts}
\newcommand{\lskip}{\vskip \baselineskip}
\newcommand{\nskip}{\lskip \noindent}
\newcommand{\hatn}{\mbox{$\bm{\hat{n}}$}}
\newcommand{\bdot}{\mbox{$\bm{\: \cdot \:}$}}
\newcommand{\bm}[1]{\mbox{\boldmath$ #1 $}}
\newcommand{\half}{\mbox{$\frac{1}{2}$}}
\newcommand{\halfskip}{\vskip 0.5\baselineskip}
\newcommand{\be}{\halfskip \begin{equation}}
\newcommand{\ee}{\end{equation} \halfskip \noindent}
\newcommand{\ba}{\halfskip \begin{eqnarray}}
\newcommand{\ea}{\end{eqnarray} \halfskip \noindent}
\newcommand{\qbar}{\bar{q}}

\title[Numerical scheme for diffusive shock acceleration]{A more accurate numerical scheme for diffusive shock acceleration}
\author[Achterberg \& Schure]{A. Achterberg \& K.M. Schure\\
Astronomical Institute, Utrecht University, Princetonplein 5, 3584 CC Utrecht, The Netherlands} 
\begin{document}
\date{Accepted... , Received...}

\maketitle

\begin{abstract}
We present a more accurate numerical scheme for the calculation of diffusive shock acceleration of cosmic rays using Stochastic Differential Equations.
The accuracy of this scheme is demonstrated using a simple analytical flow profile that contains a shock of finite width and
a varying diffusivity of the cosmic rays, where the diffusivity decreases across the shock. We compare the results for the
slope of the momentum distribution with those obtained from a perturbation analysis valid for finite but small shock width.
These calculations show that this scheme, although computationally more expensive, provides a significantly better performance than the Cauchy-Euler type schemes 
that were proposed earlier in the case where steep gradients in the cosmic ray diffusivity occur. For constant diffusivity the proposed scheme gives similar
results as the Cauchy-Euler scheme.
\end{abstract}


\section{Introduction}

One of the leading candidates for the mechanism responsible for the acceleration of Galactic cosmic rays is diffusive shock acceleration (DSA).
Proposed originally by Krimsky (1977), Axford, Leer \& Skadron (1977), Bell (1978a,b) and Blandford \& Ostriker (1978), the theory is now
well-understood in the case of non-relativistic shocks (shock speed $V_{\rm s} \ll c$), see for instance the reviews of Drury (1983),
Blandford \& Eichler (1987), Achterberg (2001) and Malkov \& Drury (2001).  In the test-particle limit in a steady flow, where the accelerated particles have no influence on the flow,
DSA at a strong, infinitely thin hydrodynamic shock yields a power-law distribution in momentum for the accelerated particles with a spectral 
index $q = - {\rm d} \ln n(p)/{\rm d} \ln p \simeq 2$, where $p$
is the particle momentum and $n(p)$ the momentum distribution. 
Such a power law is inferred for cosmic rays at the source between several GeV/nucleon and $\sim 100$ TeV per nucleon, after a correction of the
spectrum observed at Earth for the effects of propagation inside (and escape from) the Galaxy. 
In time-dependent flows with a complex flow geometry, in flows that contain multiple shocks and in particular when the cosmic rays through their pressure 
significantly decelerate the pre-shock flow, analytical results are more difficult to obtain. In those cases one often has to resort to numerical methods to solve the basic equations. 

When simulating DSA
one essentially tries to determine the cosmic ray density $N(\bm{x} \: , \: p \: , \: t)$
in reduced phase space $(\bm{x} \: , \: p = |\bm{p}|)$ as a
function of time. Two main approaches are possible: one either solves the Vlasov equation for the distribution function $N(\bm{x} \: , \: p \: , t)$
directly, or one uses an algorithm that constructs  $N(\bm{x} \: , \: p \: , \: t)$ from particle positions in phase space that have been obtained
by direct simulation of representative particle orbits. The first approach requires the solution of a partial differential equation in (reduced) phase space, which is
generally computationally expensive as it requires matrix inversion in schemes such as the Crank-Nicholson method (e.g. Potter, 1973) that is needed
in order to stably solve a diffusion-advection type equation with sufficient accuracy. The advantage of course is that one obtains the distribution function directly. 
The second approach is simple to implement in arbitrary geometries and can use very accurate integration schemes as one in principle solves an ordinary differential equation.
Its disadvantage is that one must take measures such as particle splitting (see below) in order to minimize the effect of Poisson noise that is unavoidable as one uses
a finite number of particles to construct the distribution function.

This paper is concerned with the second method, which involves the solution of {\em stochastic differential equations} (SDEs). 
It was first proposed in this context by Achterberg \& Kr\"ulls (1992). Recent applications of this method include Marcowith \& Casse (2010) and Schure {\em et al,} (2010).

We describe a relatively simple stochastic predictor-corrector scheme for diffusive shock acceleration.  It works well 
in the presence of strong gradients in the diffusivity of the particles. In that case this scheme 
is considerably more accurate than the simple Cauchy-Euler scheme, 
which fails to return accurate results for the spectral slope $q$. The slope returned by the Cauchy-Euler scheme is consistently too steep, which indicates that
the shock transition (where the acceleration takes place) is not sampled accurately.
If the diffusivity gradient is small or vanishes, both schemes give almost identical results.

We outline the different numerical approaches in Section 2, discuss the need for a better scheme in Section 3, introduce a predictor-corrector type scheme in Section 4 and
evaluate its performance in Section 5. Conclusions are found in Section 6.

\section{Numerical approaches to DSA using stochastic differential equations}

The aim of this paper is to construct the energy distribution of particles (cosmic rays) that are accelerated in a prescribed flow containing a shock, 
thereby limiting ourselves to the test-particle case.
The particles undergo spatial diffusion with respect to the flow
as a result of pitch angle scattering on magnetic fluctuations, and gain (or lose) energy due to compression (expansion) of the fluid.
Other processes, such as radiation losses and stochastic acceleration by plasma waves, can be added in a simple manner but will be neglected here.

We consider the simplest case of a one-dimensional, steady flow along the $x$-axis with fluid velocity $V(x)$.
Let
\be
\label{Fdef}
	N(x \: , \: y \: , \: t) \equiv \frac{{\rm d} {\cal N}}{{\rm d}x \: {\rm d}y}
\ee
be the cosmic ray distribution function, where $y$ is a logarithmic momentum variable, defined formally as
\be
\label{ydef}
	y = \ln (p/mc) \; .
\ee
Here $m$ is the particle rest mass and $c$ is the velocity of light. The particles are coupled to the fluid by frequent pitch angle scattering
by scattering centers (magnetic field fluctuations due to hydromagnetic waves) that are themselves tied to the flow. In the simplest case,
where these field fluctuations are advected passively by the flow with the local flow velocity $V(x)$, 
the cosmic ray distribution $N(x \: , \: y \: , \: t)$ 
satisfies the equation (cf. Skilling, 1975):
\be
\label{Skeqn}
	\frac{\partial N}{\partial t} + \frac{\partial}{\partial x} \left( V(x) \: N - D(x) \: \frac{\partial N}{\partial x} \right) = 
	\frac{1}{3} \left( \frac{{\rm d} V}{{\rm d} x} \right) \: \frac{\partial N}{\partial y} \; .
\ee
Here $N(x \: , \: y \: , \: t)$ is the distribution (averaged over gyration phase in the laboratory frame, 
the shock rest frame in the application discussed below.
$D(x)$ is the position-dependent spatial diffusion coefficient. 
For simplicity we assume this quantity to be independent of $p$ (and $y$), but this is
not important for what follows. The limitation to a steady flow is also not fundamental: the same techniques can be used to follow cosmic rays 
in a time-varying flow.

It is known that the solutions to this Fokker-Planck type equation, which essentially expresses the propagation
of particles in the two-dimensional phase space $(x \: , \: y)$, can be constructed by following particles that satisfy a stochastic differential equation
(SDE) (e.g. Gardiner, 1983, Ch.5; {\O}ksendal, 1992, Ch. 3). 

\subsection{Spatial transport}

Spatial transport in this case is described by a SDE of the type (Achterberg \& Kr\"ulls, 1992):
\be
\label{SDE}
	{\rm d}x = U(x ) \: {\rm d}t + \sqrt{2 D(x)} \: {\rm d} W_{t} \; .
\ee
In Eqn. (\ref{SDE}) the quantity
\be
\label{Udef}
	U(x) \equiv V(x) + \frac{{\rm d} D(x)}{{\rm d} x} 
\ee
is an effective advection velocity that includes a drift term due to the  
dynamical friction that results from a spatial gradient in the diffusivity. 

The first term in
SDE (\ref{SDE}) is a deterministic term $\propto {\rm d} t$ that describes the average flow of particles 
while the second stochastic term represents the spatial diffusion. 
That last term involves a infinitesimal Wiener process ${\rm d}W_{t}$, which satisfies
\be
	\left< {\rm d} W_{t} \right> = 0 \; \; \; ,  \; \; \; \left< {\rm d} W_{t}^2 \right> = {\rm d}t \; .
\ee
The brackets represent an average over many statistically independent realizations of this Wiener process.

In practice, this average is achieved
by simulating a large number of particles with statistically independent orbits.
The distribution of these particles in phase space approaches the solution of the corresponding Fokker-Planck equation, 
provided one uses a sufficient number of particles
in order to minimize the effects of Poisson noise.

In the context of Diffusive Shock Acceleration the typical length scale in the cosmic ray precursor ahead of the shock is the diffusion length
\be
	L_{\rm diff} = D/V \; .
\ee
This implies that the drift velocity within the precursor is of order
\be
	V_{\rm drift} = \frac{{\rm d} D(x)}{{\rm d} x} \simeq \frac{\Delta D}{L_{\rm diff}} = \left( \frac{\Delta D}{D} \right) \: V \; .
\ee
Taking $\Delta D \simeq D$ and $V \simeq V_{\rm s}$, with $V_{\rm s}$ the shock speed, one finds that $V_{\rm drift} \simeq V_{\rm s}$.
Formally problems arise if $V_{\rm drift} \simeq v$ with $v$ the particle velocity. 
However, in that case the diffusion approximation that is the basis of this method fails. 
The diffusion approximation presupposes that
the anisotropy in the {\em exact} momentum distribution of the accelerated particles, which is of order $V_{\rm s}/v$, is small. 
In relativistic shocks, where $V_{\rm s} \simeq c$, this is never the case.
When $V_{\rm drift} \simeq v$ ($\simeq c$ for relativistic particles)
one has to simulate the pitch angle scattering of particles directly, as is routinely done in simulations of particle acceleration 
at relativistic shocks (e.g. Bednarz \& Ostrowski, 1998; Achterberg {\em et al.}, 2001).

Within the shock transition itself one expects a change in diffusivity $\Delta D \simeq D$ over the shock width $L_{\rm s}$ so that
\be
	V_{\rm drift} \simeq \frac{D}{L_{\rm s}} \simeq v \: \left( \frac{\lambda_{\rm mfp}}{3 L_{\rm s}} \right) \; .
\ee
Here $\lambda_{\rm mfp}$ is the scattering mean free path so that the diffusion coefficient equals $D = v \lambda_{\rm mfp}/3$.
The diffusion approximation requires $V_{\rm drift} \ll v$ so {\em formally} the method fails if $L_{\rm s} \le \lambda_{\rm mfp}$.
However, in that case one may as well approximate the shock as a discontinuity with a stepwise jump in the diffusivity and flow speed. 
A scaling method for dealing with such sudden step-like jumps in the diffusion coefficient and velocity has been devised by Zhang (2000). 
It can be applied in this case so that one solves an equivalent advection-diffusion problem where
{\em no} drift term due to the diffusivity jump at the shock occurs, and the method outlined below is valid as long as the diffusion approximation applies
to particles ahead of and behind the shock front.

\subsection{Change in particle momentum}

Transport in momentum (in the absence of radiation losses or second-order Fermi acceleration by waves) follows from
\be
\label{momch}
	{\rm d}y \simeq \frac{{\rm d} p}{p} = - \frac{1}{3} \left( \frac{{\rm d} V}{{\rm d} x} \right) \: {\rm d}t \equiv \omega(x) \: {\rm d}t \; .
\ee
This gives the momentum changes in response to compressions or rarefactions in the flow, with $\omega(x)$ the local acceleration rate.
The effect of radiation losses and of second-order Fermi acceleration can be added in a simple manner.

\subsection{The Cauchy-Euler scheme}

The simplest numerical scheme that one can use to simulate this diffusion-advection process is the explicit
Cauchy-Euler scheme (CES), see for instance Achterberg \& Kr\"ulls, 1992. This scheme
advances $x$ and $y$ in time (time step: $\Delta t$) with increments in position and log momentum $\Delta x$ and $\Delta y$ given by:
\ba
\label{CES}
	\Delta x & = &   U(x) \: \Delta t +  \sqrt{2 D(x) \Delta t}  \: \xi_{t} \equiv \Delta x_{\rm s} +  \Delta x_{\rm diff}
	\: \xi_{t} \nonumber \\
	& & \\
	 \Delta y & = &  \omega(x) \: \Delta t \; . \nonumber
\ea
Here $\xi_{t}$ is a normally distributed Wiener process with zero average and unit dispersion, in the notation of Kloeden \& Platen (1992):
\be
\label{normdist}
	\xi_{t} \in {\rm N}(0 , 1) \; .
\ee
The quantity
\be
\label{Dstep}
	\Delta x_{\rm diff} \equiv
	\sqrt{2 D(x) \Delta t} 
\ee
is the rms diffusive step, given the time increment $\Delta t$.  The statistically sharp (non-stochastic) step is (see Eqn. \ref{Udef}) 
\be
	\Delta x_{\rm s} =  V \: \Delta t  + \frac{{\rm d} D}{{\rm d} x} \: \Delta t \equiv \Delta x_{\rm adv} + \Delta x_{\rm drift} \; .
\ee
It consists of the advective step $\Delta x_{\rm adv} = V \: \Delta t$ due to the plasma flow 
and the drift term $\Delta x_{\rm drift} \propto {\rm d} D/{\rm d} x$ due to gradients in the diffusivity.
This scheme has the virtue of simplicity, and for small $\Delta x_{\rm drift}$ it can accurately describe cosmic ray acceleration near shocks with a 
thickness $L_{\rm s}$ provided the time step is chosen
in such a way that
\be
\label{acccond}
	\Delta x_{\rm adv} \ll L_{\rm s} \ll \Delta x_{\rm diff} 
\ee
(Achterberg \& Kr\"ulls, 1992), who only tested the CES for uniform diffusivity so that $\Delta x_{\rm drift} = 0$.

The first condition is necessary to resolve the shock transition, and is essentially a demand of sufficient accuracy. 
The second condition ensures that a test particle, while crossing the shock with average
velocity $U$ and before escaping into the downstream flow, crosses the shock many times in diffusive jumps that are typically a few times the shock thickness.
The simulated particles essentially mimic the behavior of a real cosmic ray undergoing acceleration near a strong shock, such as a supernova blast wave. 

The whole procedure can be thought of as a replacement of the real diffusion (with a microscopic step size) by an equivalent diffusion process with the
{\em same} diffusion coefficient but with a macroscopic step size.
The last condition then ensures that the acceleration rate $\omega(x)$
due to the compression inside the shock transition 
is sampled frequently and in a stochastic manner. In this way the simulation produces the correct spectrum of accelerated particles. 

Marcowith \& Kirk (1999) have proposed a partially
implicit scheme that uses the new particle position $x + \Delta x$ to evaluate $\Delta y$. 
They showed in the simple case of a linear shock transition and a {\em constant}
diffusivity $D$ (so that once again $\Delta x_{\rm drift} = 0$) 
that this approach leads to good results, while the condition (\ref{acccond}) can be relaxed to $\Delta x_{\rm adv} \ll \Delta x_{\rm diff}$. 
Their approach opens the possibility of using larger time steps and considering zero-thickness shocks, where the velocity change is modeled by a step function.
In our approach outlined below we will follow their method for the calculation of the momentum change in terms of $\Delta y$, but will need to 
keep an explicit algorithm for the integration of the SDE for $\Delta x$.

\section{The need for a better scheme}

During numerical experiments, where acceleration of test particles is calculated in a flow that is obtained using a numerical MHD code,
it came to our attention that the simple Cauchy-Euler scheme
does not yield satisfactory results if the drift due to gradients in the diffusivity becomes too large, in particular when
$|\Delta x_{\rm drift}| \gg |\Delta x_{\rm adv}|$.
If the scale length of the variation in the diffusion coefficient is 
$L_{\rm d} = |(1/D)(\partial D/\partial x)|^{-1}$ one has
\be
\label{dxratio}
	\left| \frac{\Delta x_{\rm drift}}{\Delta x_{\rm adv}} \right| \simeq \frac{D}{L_{\rm d} V} \equiv \frac{L_{\rm diff}}{L_{\rm d}} \; .
\ee
Here $L_{\rm diff} = D/V$ is the characteristic length of  DSA,  the {\em diffusion length} introduced above.
We have found that the CES fails to give the correct particle distribution when
\be
	L_{\rm d} \simeq L_{\rm s} \ll L_{\rm diff} = \frac{D}{V}\; .
\ee
In particular (as illustrated below) the momentum distribution of the accelerated particles produced by the CES for reasonable values of the time step $\Delta t$ 
is too steep. This implies that the average momentum gain per
shock crossing as calculated using the CES is too low and the acceleration rate $\omega(x)$ is apparently not sampled accurately by the particles following the
simulated orbits.   We reiterate that
Achterberg \& Kr\"ulls (1992) and Marcowith \& Kirk (1999) did not consider this case in their calculations: they assumed a constant diffusivity $D$. 

The case of a diffusivity that varies on a scale comparable with the shock transition is astrophysically important. 
For instance: it is often assumed that the scattering
of cosmic rays near an accelerating shock is due to saturated Alfv\'en wave turbulence where the diffusivity scales as $D \propto B^{-1}$, the case of Bohm diffusion.
In a shock in an infinitely conducting plasma the MHD shock conditions imply that $B$ increases across the shock by a factor
\be
\label{rb}
	r_{\rm B} = \sqrt{ \cos^2 \theta_{\rm B} + r^2 \: \sin^2 \theta_{\rm B}} \; .
\ee
The parameter $r = \rho_{2}/\rho_{1}$ is the shock compression ratio, and $\theta_{\rm B} = \cos^{-1}(\hatn \bdot \bm{B}_{1})$ is the inclination angle 
between the upstream magnetic field $\bm{B}_{1}$ and the normal $\hatn$ to the shock surface. Depending on the orientation of the upstream magnetic field 
one has $1 \le r_{\rm B} \le r$. Additional changes in the diffusivity can arise through the reflection and transmission of MHD waves at the shock.

Here (and in what follows) we will use the subscripts 1 (2) to describe the value of quantities ahead of (behind) the shock transition. A similar case is obtained if
the cosmic rays lead to field amplification through the Bell-Lucek instability (Bell \& Lucek, 2001): there the field is amplified on the diffusive scale $L_{\rm diff} \sim D/V$,
in addition to the amplification by compression at the shock.

The use of a fully implicit scheme is not feasible for our application, as the velocity field $V(x \: , \: t)$ and the magnetic field $B(x \: , \: t)$ 
are not analytic functions but are determined by a MHD code.
We therefore need a scheme that is explicit for the advance of the position $x$, that is: it uses the variables at time $t$ and old position $x$ 
to calculate the change $\Delta x$ and the new position $x + \Delta x$. 
Such a scheme should be sufficiently accurate, numerically stable, and should be able to deal with the drift induced by strong gradients in the diffusivity. 

For the present application it is also important to minimize the number of calculations
of the spatial derivatives (or avoid them altogether) of the velocity field and the diffusivity. 
Such derivatives tend to be noisy when they are determined from the raw output of a MHD code.
In the case of the calculation of the momentum change this can be achieved by using the method proposed by Marcowith \& Kirk (1999).  In terms of $y = \ln(p/mc)$ 
one replaces the second equation of (\ref{CES}) by:
\be
	\Delta y = - \frac{\Delta t}{3 \Delta x} \left[ V(x + \Delta x) - V(x) \: \right] \equiv \overline{\omega} \: \Delta t \; ,
\ee
assuming a steady flow for simplicity.
The position change $\Delta x$ is determined in the way outlined below. 
This algorithm essentially uses the spatial average of the acceleration rate along the orbit of a simulated particle. For a steady flow:
\be
\label{omave}
	\overline{\omega} = \frac{1}{\Delta x} \int_{x}^{x + \Delta x} {\rm d}x \:  \left[ - \frac{1}{3} \frac{{\rm d} V(x)}{{\rm d}  x} \: \right] =
	- \frac{V(x + \Delta x) - V(x) }{3 \Delta x} \; .
\ee
For time-varying flows (where ${\rm d}V/{\rm d}x \: \Longrightarrow \: \partial V/\partial x$), 
and in the unlikely case that the advective step and the diffusive step cancel each other (so that $\Delta x = 0$)
this prescription can lead to singular behaviour. 
However, this is easily caught in a numerical scheme and correctly dealt with by putting $\Delta y = 0$ in that case.

\section{A predictor-corrector scheme for spatial transport}

We have found that a scheme
proposed by Kloeden \& Platen (1992) gives excellent results in the case of strong drift due to gradients in the diffusivity, 
where the CES fails to be sufficiently accurate. It is a second-order predictor corrector scheme, called the KPPC scheme in what follows. 
In particular this scheme improves the accuracy of the spatial transport of the particles, which leads to a better sampling of the
acceleration rate $\omega(x)$ in the shock compression.
For the problem at hand this scheme takes the following form:

\subsubsection*{Step 1: first supporting position value}

As a first step one calculates a first supporting position value  $\tilde{x}$ using the Cauchy-Euler scheme:
\be
\label{s1step}
	\tilde{x} = x + U(x \: , \: t) \: \Delta t + \sqrt{2 D(x) \Delta t} \: \xi_{t}	
\ee
For simplicity we adopt a constant time step $\Delta t$. The stochastic variable 
$\xi_{t} \in {\rm N}(0 , 1)$ is drawn from a normal distribution with zero mean and unit dispersion.
Two additional supporting position values are calculated,
\be
\label{s2step}
	x_{\pm} = x + U(x \: , \: t) \: \Delta t \pm  \sqrt{2 D(x) \Delta t} \; ,
\ee
that correspond with the position of two hypothetical particles that experience $\pm$ the rms diffusive step. 
The stochastic variable $\xi_{t}$ is now {\em fixed} at the value used in (\ref{s1step}). It does not
change in any of the subsequent steps of the algorithm that are outlined below.

\subsubsection*{Step 2: predicted position value}

As a next step one calculates the predicted position $\bar{x}$ at time $t + \Delta t$ as
\be
\label{pstep}
	\bar{x} = x + \bar{U} \: \Delta t + \overline{\Delta x}_{\rm diff}(\xi_{t})  \; .
\ee
The mean velocity $\bar{U}$ used in the advective + drift term is an average velocity, defined by using the first supporting value $\tilde{x}$:
\be
\label{barU}
	\bar{U} = \half \left[ U(x) + U(\tilde{x}) \: \right] \; .
\ee
The improved stochastic diffusive step $\overline{\Delta x}_{\rm diff}(\xi_{t})$ equals
\ba
\label{barDiff}
	\overline{\Delta x}_{\rm diff}( \xi_{t}) & = 	& \left( \frac{\Delta x_{\rm diff}^{+} + \Delta x_{\rm diff}^{-} + 2 \Delta x_{\rm diff}}{4} \right) \: \xi_{t} \nonumber \\
	& & \\
	& + & \left( \frac{\Delta x_{\rm diff}^{+} - \Delta x_{\rm diff}^{-}}{4} \right) \: \left( \xi_{t}^2 - 1 \right) \; . \nonumber
\ea
Here $\Delta x_{\rm diff}$ is the rms diffusive step (\ref{Dstep}) at the old position and $\Delta x_{\rm diff}^{\pm}$ corresponds to the rms diffusive step evaluated
at the two supporting positions $x_{\pm}$ that were defined in relation (\ref{s2step}):
\be
	\Delta x_{\rm diff}^{\pm} \equiv \sqrt{2 D(x_{\pm}) \Delta t} \; .
\ee
The second term of the corrected diffusive step (\ref{barDiff}) corrects for the
`lopsidedness' of the random walk that results from the gradient in the diffusivity. 

\subsubsection*{Step 3: corrector step and final position}

One finally obtains the corrected (and final) position $x_{\rm c} \equiv x(t + \Delta t)$ at time $t + \Delta t$ in the following way:
\be
\label{cstep}
	x_{\rm c} = x + \half \left[ U(x) + U(\bar{x}) \: \right] \: \Delta t + \overline{\Delta x}_{\rm diff}(\xi_{t})  \; .
\ee
This last step uses the same diffusive step as in the predictor cycle but corrects the advective step using the predicted position $\bar{x}$ (see Eqn. \ref{pstep}). 
This is a reasonable approach as (on average) $\Delta x_{\rm diff} \gg \Delta x_{\rm adv}$.
It was already noted by Marcowith \& Kirk (1999) that a careful 
treatment of the advective step (including drift) is more important for the accuracy of the scheme than the treatment of the diffusive step. 
This scheme and the tests presented below bear that out.

A few remarks about the implementation of this scheme are in order. 

First of all, this scheme is computationally about 6 times more expensive than the Cauchy-Euler scheme.
An order-of-magnitude increase of the computational effort is typical when switching from an explicit, first-order scheme to a second-order accurate
predictor-corrector 
scheme or the closely related Runge-Kutta type schemes.

Secondly: for the term that corrects for diffusivity gradients to be effective one should {\bf not} employ an often-used
numerical approximation for the Wiener process that replaces the normal distribution for $\xi_{t}$ by a two-point distribution of values, choosing $\xi_{t} = \pm 1$, 
where the two possible signs are drawn randomly with
equal probability ${\cal P}_{+} = {\cal P}_{-} = \half$. In that numerical approximation for the Wiener process 
the second term in the expression (\ref{barDiff}) vanishes identically, and much of the scheme's
improved accuracy with respect to the Cauchy-Euler scheme is lost. 
In that respect one might expect that a symmetric three-value scheme for $\xi_{t}$, for example 
(in the notation $[value \: | \: probability]$)
\be 
\label{threepoint1}
	\xi_{t} \in \left[ - \sqrt{3} \: | \:  \frac{1}{6} \: \right] \; \; , \; \; \left[ 0 \: | \: \frac{2}{3} \: \right] \; \; , \; \; \left[ + \sqrt{3} \: | \:  \frac{1}{6} \: \right] \; , 
\ee
works better.

\section{Tests of the algorithm}

\subsection{Basic assumptions}

We have tested the KPPC scheme as described here, comparing its performance to the performance of the simpler Cauchy-Euler scheme. 
For this test we use scaled (dimensionless)
variables where the fluid velocity $V$ is measured in units of the shock speed and position $x$ along the shock normal is in units of the shock thickness $L_{\rm s}$
For clarity we keep $L_{\rm s}$ in the equations even though $L_{\rm s} = 1$ in the numerical implementation. 
The velocity $V(x)$ is in the direction of positive $x$, given by
\be
\label{Vprofile}
	V(x) = \frac{r + 1}{2r} - \frac{r - 1}{2r} \: {\rm tanh}\left( \frac{x}{L_{\rm s}} \right)  \; .
\ee
The velocity decreases with increasing $x$, from $V(-\infty) \equiv V_{1} = 1$ to $V(+ \infty) \equiv V_{2} = 1/r$. This means
that we work in the rest frame of the shock and measure the flow velocity in units of the shock velocity with respect of the upstream medium.
Here $r > 1$ is the compression ratio of the shock transition in the sense that (for this one-dimensional
steady flow) the conservation of mass implies $\rho V = {\rm constant}$, with $\rho$ the mass density. The density
contrast between the far upstream and far downstream state follows as
\be
	\frac{\rho(\infty)}{\rho(- \infty)} \equiv \frac{\rho_{2}}{\rho_{1}} = \frac{V_{1}}{V_{2}} = r \; . 
\ee 
This velocity profile models the shock as a stationary and smooth transition, with a width (velocity gradient scale) $L_{\rm s}$.  
To model a varying diffusion coefficient we adopt a diffusion coefficient that varies with position $x$ as
\be
\label{Dprofile}
	D(x) = D_{1} \: \left[ \frac{\sigma + 1}{2 \sigma} - \frac{\sigma - 1}{2 \sigma} \: {\rm tanh} \left( \frac{x}{L_{\rm d}} \right)  \: \right] \; .
\ee
Here $D_{1}$ is a constant dimensionless diffusivity that is related to the physical diffusivity $D_{\rm phys}$ far ahead of the shock
by $D_{1} = D_{\rm phys}/L_{s} V_{s}$ with
$V_{s}$ the shock velocity. The diffusivity decreases if one moves from upstream ($x < 0$) to downstream ($x > 0$) 
across the shock, with a ratio of asymptotic values equal to
\be
	\frac{D(- \infty)}{D(\infty)} \equiv \frac{D_{1}}{D_{2}} = \sigma \ge 1 \; , 
\ee
the kind of behavior one expects in astrophysical applications.
The scale length for the variation of the diffusion coefficient in these units is of order $L_{\rm d}$. 
For future use we define the quantity
\be
	\varepsilon \equiv \frac{L_{\rm s}}{L_{\rm diff}(- \infty)} = \frac{V_{1} L_{\rm s}}{D_{1}} \; .
\ee
This is essentially the {\em P\'eclet number} of the shock based on the cosmic ray diffusivity. In terms of this quantity one has
\be
\label{driftratio}
	\left| \frac{\Delta x_{\rm drift}}{\Delta x_{\rm adv}} \right| \simeq  \frac{\sigma - 1}{2 \sigma } \: \frac{L_{\rm s}}{\varepsilon L_{\rm d}} \; .
\ee
A sharp shock in the present context corresponds to $\varepsilon \ll 1$.

The main test of the algorithm lies in its ability to reproduce the spectrum predicted by the analytical theory of DSA  at a 
steady shock with $\varepsilon \ll 1$, $\sigma \simeq r$ and $L_{\rm s} \simeq L_{\rm d}$. In the limit of a infinitely thin shock with $\varepsilon =0$
(in physical terms: a shock thickness that is much smaller 
than the scattering mean free path of the accelerating cosmic rays) and
in the absence of radiation losses the predicted shape of the spectrum is a power law in momentum, with an index $q$ 
that depends only on the compression ratio $r$ (e.g. Axford, Leer \& Skadron, 1977, 
Bell, 1978; Blandford \& Ostriker, 1978). In present notation, using the momentum $p$ rather than $y = \ln(p/mc)$:
\be
\label{DSAlaw}
	N(x \: , \: p) = \frac{{\rm d} {\cal N}}{{\rm d}x \: {\rm d} \ln p} \propto p^{-q} \; \; , \; \; q(\varepsilon = 0) = \frac{3}{r - 1} \; .
\ee
In our tests of the algorithm we have assumed a diffusion coefficient that is independent of particle momentum, so 
this power-law behavior is valid uniformly across the grid, with only
the concentration of test particles varying with position $x$.

\subsection{Effect of finite shock thickness}

Since we assume a finite shock thickness, a situation typical of shocks obtained through numerical simulation, we need an expression for the slope
$q$ for finite $\varepsilon$. We use a perturbation analysis adapted from Drury (1983) and the closely related method of Schneider \& Kirk (1987).
The analysis presented below is valid when there is a small parameter $\varepsilon$, in this case the ratio of the shock thickness and the cosmic ray diffusion length:
\be
\label{epscond}
	\varepsilon \equiv \frac{L_{\rm s}}{L_{\rm diff}} \ll 1 \; .
\ee 
Consider the steady-state transport equation (\ref{Skeqn}) reformulated in terms of the 
Vlasov distribution $f(\bm{x} \: , \: p) \equiv {\rm d}{\cal N}/({\rm d}^3 \bm{x} \: {\rm d}^3 \bm{p})$.
In our application we have $f(x \: , \: p) = N(x \: , \: p)/4 \pi p^3$.
Assuming a one-dimensional steady flow in the $x$-direction the equation for $f(x \: , \: p)$ with $\partial f/\partial t = 0$ reads:
\be
\label{Skeqn2}
	V \: \frac{\partial f}{\partial x} - \frac{\partial}{\partial x} \left( D \frac{\partial f}{\partial x} \right) = 
	\frac{1}{3}\frac{{\rm d} V}{{\rm d} x} \: \left( p \frac{\partial f}{\partial p} \right) \; .
\ee
Schneider \& Kirk use a variant of the Ricatti transformation (e.g. Polyanis \& Manzhirov, 2007, Ch. 12.2), which we slightly modify here.
We introduce a dimensionless position variable $X$,
\be
	X(x) \equiv \int_{0}^{x} \: \frac{V(x') {\rm d} x'}{D(x')} \; ,
\ee 
and define (c.f. Schneider \& Kirk, 1987)
\be
	G(x \: , \: p) = -\frac{\displaystyle D \: \frac{\partial f}{\partial x}}{f(x \: , \: p)} = -\frac{\displaystyle V \: \frac{\partial f}{\partial X}}{f(x \: , \: p)}\; .
\ee
Schneider \& Kirk also assume a power-law momentum dependence,
\be
\label{powerl}
	f( x \: , \: p) \propto p^{-\qbar} \; ,
\ee
which can only be strictly justified if the diffusion coefficient is independent of momentum, the case we consider here, and for momenta well above the injection momentum.
Note that we have $\qbar = q + 3$.
In that case $G(x \: , \: p)$ is a function $G(X)$ of position alone.
Equation (\ref{Skeqn2}) can be written as a non-linear ordinary differential equation for $G(X)$: 
\be
\label{Reqn}
		\frac{{\rm d}G}{{\rm d}X} + \frac{\qbar}{3} \: \frac{{\rm d}V}{{\rm d}X} = G + \frac{G^2}{V} \; .
\ee
This is the relation derived by Schneider \& Kirk (1987), generalized to the case of a position-dependent diffusion coefficient.
The boundary conditions for $G(X)$ at $X = \pm \infty$ are
\be
\label{BCs}
	G(- \infty) = - V(-\infty) \equiv - V_{1} \; \; , \; \; G(+ \infty) = 0 \; . 
\ee
The first condition assumes that there are no pre-existing particles far ahead of the shock so that asymptotically $f(x \: , \: p) \propto {\rm exp}(X)$ \
for large negative $X$, where $V(X) \simeq V_{1}$ is approximately constant.
The second condition, which states that the diffusive contribution $\propto \partial f/\partial x$
to the flux vanishes asymptotically far behind the shock, ensures
that the particle density remains finite as $X \: \longrightarrow + \infty$. 

We now assume that condition (\ref{epscond}) is satisfied, meaning that the shock transition must be sufficiently sharp.
In that case, the left-hand side of Eqn. (\ref{Reqn}) will be large in a region $\Delta X \sim \varepsilon$ when compared with the two terms on the right-hand side.
This behavior can be formalized by using $\varepsilon$ as a formal ordering parameter, replacing (\ref{Reqn}) by
\be
\label{Reqn2}
	\frac{1}{\varepsilon} \left( \frac{{\rm d}G}{{\rm d}X} + \frac{\qbar}{3} \: \frac{{\rm d}V}{{\rm d}X} \right) = G + \frac{G^2}{V} \; .
\ee 
We seek solutions of the form
\be
\label{Gexp}
	G(X) = G_{0}(X) + \varepsilon \: G_{1}(X) + \varepsilon^2 \: G_{2}(X) + \cdots
\ee
and expand the slope as
\be
\label{qexp}
 	\qbar = \qbar_{0} + \varepsilon \: \qbar_{1} + \varepsilon^2 \: \qbar_{2} + \cdots 
\ee
We can now solve (\ref{Reqn2}) at each order of $\varepsilon$, putting $\varepsilon = 1$ at the end of the calculation.
At leading order ($\varepsilon^{-1}$) one has
\be
\label{zeroG}
	\frac{{\rm d} G_{0}}{{\rm d}X} + \frac{\qbar_{0}}{3} \: \frac{{\rm d}V}{{\rm d}X} = 0 \; ,
\ee
subject to the boundary conditions (\ref{BCs}) for $G_{0}(X)$: 
\be
G_{0} (- \infty) = - V_{1} \; \; \mbox{and} \; \;  G_{0}(+ \infty) = 0 \; . 
\ee
The solution is elementary:
\be
\label{Gzero}
	G_{0}(x) = \frac{\qbar_{0}}{3} \: \left[ \: V_{2} - V(X) \: \right] \; ,
\ee
with $V_{2} = V(+ \infty)$ the asymptotic velocity downstream.
The zero-order slope $\qbar_{0}$ can be found by integrating (\ref{zeroG}) from $X = - \infty$ to $X = + \infty$ and using the boundary conditions.
Another elementary calculation gives
\be
	\frac{\qbar_{0}}{3} \: \left(V_{2} - V_{1} \right) + V_{1} = 0 \; .
\ee
One finds (as expected) that $\qbar_{0}$ is the slope associated with an infinitely thin shock, where the velocity jumps from $V_{1}$ to $V_{2} < V_{1}$ at $x = 0$:
\be
\label{qzero}
	\qbar_{0} = \qbar(\varepsilon = 0) = \frac{3 V_{1}}{V_{1} - V_{2}} \; . 
\ee
This also ensures the the boundary condition at $X = - \infty$ is satisfied as (\ref{qzero}) substituted into (\ref{Gzero}) 
implies
\be
\label{G0sol}
	G_{0}(X) = - \frac{V_{1} \: \left( V(X) - V_{2} \right)}{V_{1} - V_{2}} \; . 
\ee
At next order ($\varepsilon^{0}$) one has:
\be
\label{firstG}
	\frac{{\rm d} G_{1}}{{\rm d}X} + \frac{\qbar_{1}}{3} \: \frac{{\rm d}V}{{\rm d}X} = G_{0} + \frac{G_{0}^2}{V}  \; ,
\ee
subject to the boundary condition
\be
\label{BC1}
	G_{1}(- \infty) = G_{1}(+ \infty) = 0 \; .
\ee
Integrating (\ref{firstG}) from $X = - \infty$ to $X = +\infty$ immediately yields a relation for $\qbar_{1}$:
\be
	\frac{\qbar_{1}}{3} \: \left( V_{2} - V_{1} \right) = \int_{- \infty}^{+ \infty} {\rm d}X \: \left(G_{0} + \frac{G_{0}^2}{V} \right) \; .
\ee
Solving for $\qbar_{1}$, using (\ref{qzero}) and (\ref{G0sol}):
\be
\label{qone}
	\qbar_{1} = \qbar_{0} \:  \int_{- \infty}^{+ \infty} {\rm d}X \: 
	\frac{\displaystyle V_{2} \left(V_{1} - V \right) \left(V - V_{2} \right)}
	{\displaystyle V \: \left(V_{1} - V_{2} \right)^2} \; .
\ee
The function $G_{1}(X)$ is
\ba
\label{G1}
	G_{1}(X) & = &  \frac{\qbar_{1}}{3} \: \left( V_{1} - V \right) \nonumber \\
	& & \\
	& & \; \; \; \; \; - \int_{- \infty}^{X} {\rm d}X' \: 
	\frac{\displaystyle V_{1} V_{2} \left(V_{1} - V' \right) \left(V' - V_{2} \right)}
	{\displaystyle V' \: \left(V_{1} - V_{2} \right)^2} \; . \nonumber
\ea
Here $V' \equiv V(X')$. Note that $\qbar_{1}$ and $G_{1}(X)$ vanish automatically if one uses the step function velocity profile of an infinitely thin shock, 
with $V(X) = V_{1}$ for $X < 0$ and $V(X) = V_{2}$ for $X \ge 0$.
\nskip
At order $\varepsilon$ one finds the following equation for $G_{2}(X)$:
\be
\label{secondG}
	\frac{{\rm d} G_{2}}{{\rm d}X} + \frac{\qbar_{2}}{3} \: \frac{{\rm d}V}{{\rm d}X} = G_{1} \: \left(1 + \frac{2 G_{0}}{V} \right) \; ,
\ee 
subject to the boundary condition $G_{2}( - \infty) = G_{2}(+ \infty) = 0$. Integrating (\ref{secondG}) from $X = - \infty$ to $X = +\infty$ 
yields an equation for $\qbar_{2}$:

\ba
\label{qtwo}
	\qbar_{2} & = &  - \frac{3}{V_{1} - V_{2}} \int_{- \infty}^{+ \infty} {\rm d}X \: G_{1}(X) \: \left(1 + \frac{2 G_{0}(X)}{V(X)} \right) 
	\nonumber \\
	& & \\
	& = & 
	\frac{3}{V_{1} - V_{2}} \int_{- \infty}^{+ \infty} {\rm d}X \: G_{1}(X) \: 
	\left(     
	\frac{\left( V_{1} + V_{2} \right)V(X) - 2V_{1} V_{2}}{V (X)\left( V_{1} - V_{2} \right)} \right)\; . 
	\nonumber 
\ea
This procedure can be extended to higher order, but little is gained at the expense of increasingly complex mathematics.

\subsubsection{Specific examples}

We use two examples of immediate importance for a test of the numerical scheme advocated here. Table 1 gives the parameters as used in the
numerical simulations presented below.

\subsubsection*{Model 1: hyperbolic tangent velocity profile and constant diffusivity}

The first example is the case of a uniform diffusivity $D(x) = D_{1}$, which formally corresponds to $\sigma = 1$ and $L_{\rm d} = \infty$.
This is the case where the CES is known to yield good results. This case has been treated before by Axford, Drury \& Summers (1982), who show that for
the hyperbolic tangent velocity profile (\ref{Vprofile}) adopted here the cosmic ray transport equation can be solved analytically. 
They find an asymptotic slope of the momentum distribution equal to 
\be
\label{DASq}
	\qbar = \qbar_{0} \: \left( 1 + \frac{V_{2} L_{\rm s}}{2 D_{1}} \right) \; .
\ee
Here $\qbar_{0} = 3r/(r - 1)$. 
The perturbation expansion used here (and in a slightly different form by Drury (1983)) reproduces this result. The hyperbolic tangent velocity law (\ref{Vprofile})
implies
\be
\label{Vder}
	\frac{{\rm d}V}{{\rm d}x} = - \frac{2}{L_{\rm s}} \: \frac{\left( V_{1} - V \right) \left( V - V_{2} \right)}{V_{1} - V_{2}} \; .
\ee 
Substituting this into the generally valid expression (\ref{qone}), together with ${\rm d}X = V \: {\rm d}x/D_{1}$, one finds:
\be
	\qbar_{1} = - \frac{\qbar_{0} V_{2} L_{\rm s} }{2 D_{1} (V_{1} - V_{2}) } \: \int_{-\infty}^{\infty} {\rm d} x \: \left( \frac{{\rm d}V}{{\rm d} x} \right)  = 
	\qbar_{0} \: \frac{V_{2} L_{\rm s}}{2 D_{1}} \; .
\ee 
This agrees with result (\ref{DASq}) of Drury et al (1982). Using this in relation (\ref{G1}) one finds that $G_{1}(X) = 0$, which implies
$G_{n} = q_{n} = 0$ for $n \ge 2$. Here the perturbation expansion breaks off at order $\varepsilon$ and yields the {\em exact} asymptotic 
result, as noted before by Drury (1983). In terms of the compression ratio $r = V_{1}/V_{2}$, the diffusion length far upstream $L_{\rm diff}^{-\infty} = D_{1}/V_{1}$ and
$q = \qbar - 3$ one has:
\be
\label{qDAS}
	q = \frac{3}{r - 1} \: \left( 1 + \frac{V_{1} L_{\rm s}}{2 D_{1}} \right) = \frac{3}{r - 1} \: \left(1 + \frac{\varepsilon}{2} \right) \; .
\ee 

\subsubsection*{Model 2: hyperbolic tangent profile and constant diffusion length}

As a second example we consider the case of a constant diffusion length:
\be
	L_{\rm diff} = \frac{D(x)}{V(x)} = \frac{D_{1}}{V_{1}}\; .
\ee
This example is important as a test case of the predictor-corrector algorithm used here as it has a strong gradient in the diffusivity.
Formally it corresponds to $\sigma = r$ and $L_{\rm d} = L_{\rm s}$ so that in the shock
\be
	\left| \frac{\Delta x_{\rm drift}}{\Delta x_{\rm adv}} \right| \simeq \frac{r - 1}{2 r \varepsilon} \; ,
\ee
which becomes large if $\varepsilon \ll 1$ for thin shocks.
If one adopts the hyperbolic tangent profile (\ref{Vprofile})/(\ref{Vder}) and uses the fact that
\be
	{\rm d}X = \frac{{\rm d}x}{L_{\rm diff}} \; ,
\ee
relation (\ref{qone}) yields:
\ba
	\qbar_{1} & = &  - \: \frac{\qbar_{0} L_{\rm s}}{2 L_{\rm diff}} \: \left( \frac{V_{2}}{V_{1} - V_{2}} \right) \int_{- \infty}^{\infty} {\rm d} x \: 
	\left( \frac{1}{V} \frac{{\rm d}V}{{\rm d}x} \right) \nonumber \\
	& & \\
	& = & 
	\frac{\qbar_{0} L_{\rm s}}{2 L_{\rm diff}} \: \left( \frac{V_{2}}{V_{1} - V_{2}} \right) \: \ln \left( \frac{V_{1}}{V_{2}} \right) \; . \nonumber
\ea
The function $G_{1}$ can be calculated, but the integral over $G_{1}$ that determines the next order correction $\qbar_{2}$ to the slope can not be expressed in
elementary functions.  Limiting ourselves to the first-order correction one has in terms of $q = \qbar - 3$ and $r = V_{1}/V_{2}$:
\be
	q \simeq \frac{3}{r - 1} \:  \left(1 + \frac{ r \: (\ln r) \: L_{\rm s}}{2(r - 1) L_{\rm diff}} \right) = 
	\frac{3}{r - 1} \:  \left(1 + \frac{ r \: (\ln r) \: \varepsilon}{2(r - 1)} \right) \; .
\ee
Here the steepening due to a finite shock width $\sim L_{\rm s}$ is more pronounced than in the previous cases as the diffusion length near the shock transition is smaller
compared to the first case. 
For instance: in a shock with compression ratio $r= 4$, the value expected for a strong shock in a mono-atomic gas, one has
$r \: \ln r/2(r - 1) \simeq 0.924$. We have obtained the second-order term for the important case $r = 4$ through numerical integration 
of the integral in the expression for $\qbar_{2}$. We find:
\ba
\label{ris4}
	q(r = 4) & = &  1 + 0.924 \: \frac{L_{\rm s}}{L_{\rm diff}} + 0.095 \: \frac{L_{\rm s}^2}{L_{\rm diff}^2} \nonumber \\
	& & \\
	& = & 
	 1 + 0.924 \: \varepsilon + 0.095 \: \varepsilon^2 \; . \nonumber
\ea
Note that the end result in both cases is a series in $\varepsilon = L_{\rm s}/L_{\rm diff}$. 
It should be pointed out that the term $\propto \varepsilon^2$ is small in the
second case, and vanishes completely in the first case. The smallness of the second-order correction to $q$
seems to be a rather general property of this expansion in $\varepsilon$ for reasonable velocity profiles. 
As a further example: in the case of a linear velocity profile, where
\be
\label{linprofile}
	V(x) = \left\{ \begin{array}{ll}
	V_{1} & \mbox{for $x < - L_{\rm s}/2$,} \\
	& \\
	{\displaystyle \frac{V_{1} + V_{2}}{2} - \frac{V_{1} - V_{2}}{L_{\rm s}}} \: x & \mbox{for $-L_{\rm s}/2 \le x \le L_{\rm s}/2$,} \\
	& \\
	V_{2} & \mbox{for $x > L_{\rm s}/2$,} \\
	\end{array} 
	\right. 
\ee
and for a constant diffusion coefficient $D = D_{1}$ ($\sigma = 1$, $L_{\rm d} = \infty$), the same procedure yields:
\be
	q = \qbar - 3 \simeq \frac{3}{r - 1} \left( 1 + \frac{\varepsilon}{6} + \frac{(r +1) \varepsilon^2}{360 r} \right) \; .
\ee
For $r = 4$ this is
\be
	q(r = 4) = 1 + \frac{\varepsilon}{6} + \frac{\varepsilon^2}{288} \; .
\ee
These three examples suggest that the results obtained here for the slope $q(\varepsilon)$ are applicable even for $\varepsilon \simeq 1$. 
The simulations presented in the next Section bear this out.

\begin{table}
\caption{Model parameters}
\label{table1}
\centering
\begin{tabular}{c c c c c} \hline \hline
& & & & \\
{\bf Model} & $r$ & $\sigma$ & $L_{\rm d}/L_{\rm s}$ & $V_{1} \: \Delta t/L_{\rm s}$\\
& & & & \\ \hline
& & & & \\
1 & 4 & 1 & $\infty$ & 0.05\\
& & & & \\
2 & 4 & 4 & 1 & 0.05\\
& & & & \\ \hline \hline
\end{tabular}
\end{table}

\subsection{Numerical results}

The two figures below show the results of numerical simulations for the two cases listed in Table 1. The approach is similar to the one employed by
Achterberg and Kr\"ulls (1992): particles are injected close to the shock, and followed over a grid that extends from $X = - 10$ to $X = + 10$.
Particles are detected as they cross the downstream boundary $X_{\rm max} \simeq V_{2} x_{\rm max}/D_{2} = 10$, which acts as an absorber. 
The influence of a downstream absorbing boundary
on the slope $q$ decays as ${\rm exp}(-X_{\rm max})$ with respect to unity, and is negligibly small for these parameters. 
Particle splitting is used at intervals equidistant in 
$\log p$ to minimize the effects of Poisson noise at large momenta where fewer particles reside in the distribution. 
The spectra obtained in this way are strict power laws that extend over five
decades ($\Delta y \simeq 11.5$) in particle momentum. We use a fixed time step that corresponds to $V_{1} \: \Delta t = 0.05 \: L_{\rm s}$, 
so that the advective step resolves the shock transition.
In practice, good results are obtained if $\Delta x_{\rm adv} \lse 0.1 \: L_{\rm s}$.
The diffusion coefficient
varies from $D_{1} = 1$ to $D_{1} = 100$, which corresponds to a diffusive step in the range $\Delta x_{\rm diff} \simeq 0.3 - 3$. Note that
in our implementation we have scaled the spatial coordinate $x$ with the shock width so that $L_{\rm s} = 1$. 

\begin{figure}
	\includegraphics[width=84mm]{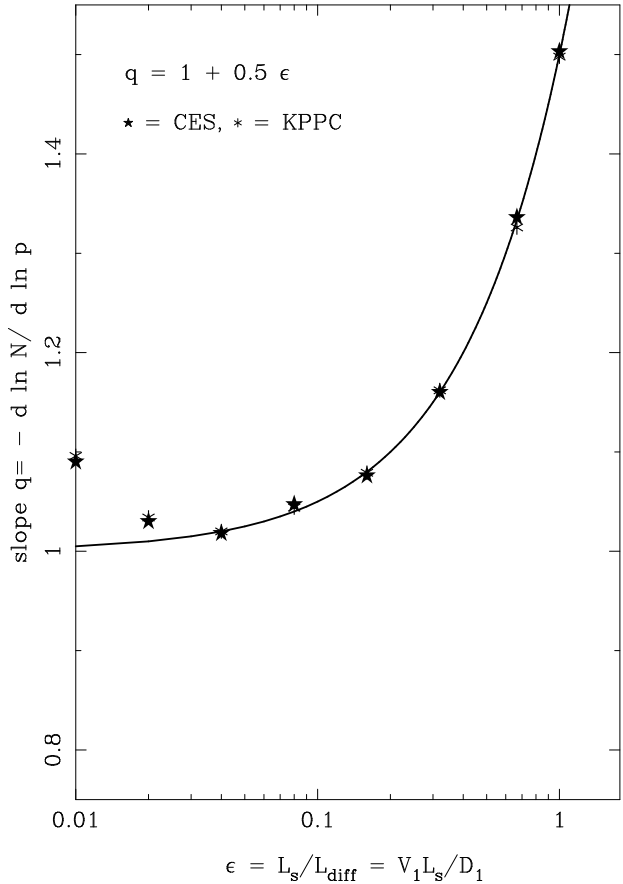}
	\caption[]{Results for Model 1: the case of a hyperbolic tangent velocity profile and constant diffusivity.
	The solid curve gives the result (\ref{qDAS}), the open stars are the results of the KPPC scheme and the solid stars
	the results obtained using the CES. The shock compression ratio equals $r = 4$, the case of a strong hydrodynamical shock.
	The parameter $\varepsilon = L_{\rm s}/L_{\rm diff}$ varies from 0.01 to unity.
	Note that the figure employs a logarithmic scale for $\varepsilon$.}
\label{fig1}
\end{figure}

\begin{figure}
	\includegraphics[width=84mm]{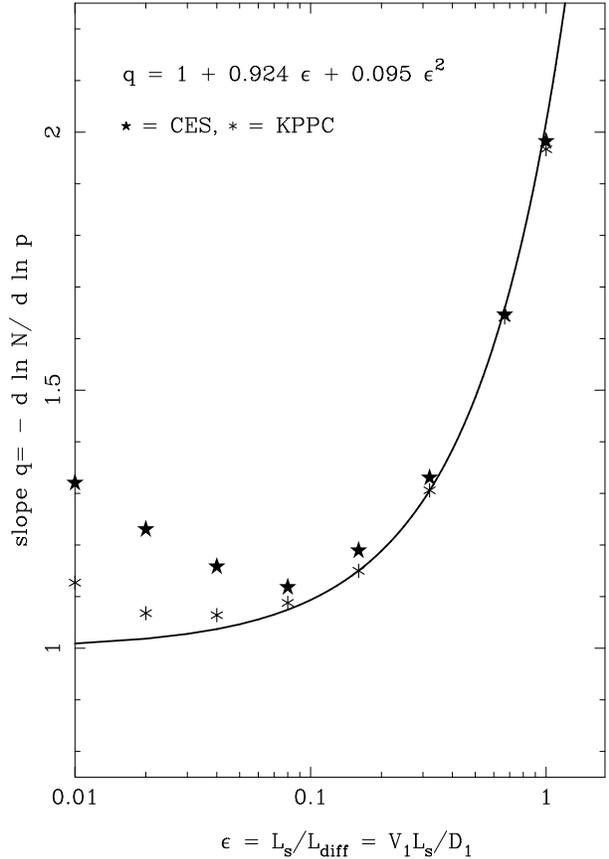}
	\caption[]{Results for Model 2: the case of a hyperbolic tangent velocity profile and constant diffusion length $L_{\rm diff} = D(x)/V(x)$.
	The solid curve gives the result (\ref{ris4}), the open stars are the results of the KPPC scheme and the solid stars
	the results obtained using the CES. The shock compression ratio equals $r = 4$, the case of a strong hydrodynamical shock.
	Again a log-linear scale is employed in this figure. Note that the vertical scale differs from the one used in Figure 1.}
\label{fig2}
\end{figure}

Figure 1 shows the results obtained for the case of a constant diffusion coefficient (Model 1). In this model there is no drift term.
In this case the CES and the KPPC scheme give comparable results that closely follow the theoretical prediction up to $\varepsilon = L_{\rm s}/L_{\rm diff} = 0.04$.
For smaller values of $\varepsilon$ (i.e. larger values of $D_{1}$) the both schemes become inaccurate as they under-sample the acceleration rate $\omega(x)$ in the shock transition. 

Figure 2 shows the results for Model 2, where $L_{\rm diff} = D(x)/V(x)$ is kept constant so that $L_{\rm s} = L_{\rm d}$ and $\sigma = r = 4$. 
This is the model with a large drift in the shock due to the
gradient in the diffusivity. The results for the slope of the momentum distribution obtained using the CES (the filled stars) deviate significantly from the slope 
obtained using the perturbation expansion if $\varepsilon = L_{\rm s}/L_{\rm diff} = L_{\rm d}/L_{\rm diff} < 0.1$. In contrast, the KPPC scheme gives significantly 
better results that are usable up to $\varepsilon \simeq 0.02$. The error in the value of the slope $q$ returned by the KPPC scheme is typically 3 times smaller than
the error produced by the CES. 
The deviation from the analytical result becomes significant if the magnitude of the drift term in the statistically sharp spatial 
step $\Delta x_{\rm s}$ becomes of the same order as the shock thickness:
\be
	\left| \Delta x_{\rm drift} \right| = \left| \frac{{\rm d} D}{{\rm d}x}  \right| \: \Delta t \ge L_{\rm s} \; .
\ee
Making the estimate $|{\rm d}D/{\rm d}x| \simeq |\Delta D|/L_{\rm s}$ as it applies to typical situations where the diffusion coefficient jumps by an amount
$\Delta D$ across the shock, the accuracy of the KPPC scheme is lost if
\be
	\frac{\Delta x_{\rm drift}}{L_{\rm s}} = \frac{|\Delta D|}{D_{1}} \frac{V_{1} \Delta t}{\varepsilon L_{\rm s}} > 1 \; .
\ee
For instance: in the results shown in Figure 2 the deviation in the slope returned by the KPPC scheme becomes large when $|\Delta x_{\rm drift}| \simeq 2 L_{\rm s}$.

In our test of the KPPD scheme we have assumed that the diffusion coefficient is independent of momentum. In many astrophysical applications one expects the mean free
path to increase with momentum so that the diffusion coefficient scales as $D \propto p^{\alpha}$. For instance: one often assumes {\em Bohm Diffusion} with a mean-free-path
equal to the gyro radius, $\lambda_{\rm mfp} = r_{\rm g} \simeq pc/qB$ where $q$ is the particle charge. For relativistic particles ($v\sim c$) this implies $D \propto p$.
The KPPD scheme should be able to give reliable result in this case also, as long as the time steps are such that $|\Delta x_{\rm drift}| \lse L_{\rm s}$.
The scaling $D \propto p^{\alpha}$ implies $\Delta x_{\rm drift} \propto p^{\alpha}$, so it may be necessary, depending on the dynamic range in $p$, 
to employ the scheme with smaller time steps for the high-energy particles
in the population than for the low-energy particles.

\section{Discussion and conclusions}

We have argued that the simulation of diffusive shock acceleration of cosmic rays through the numerical integration of stochastic differential equations
needs a more accurate, second-order scheme when large gradients in the cosmic ray diffusivity arise. This situation is astrophysically relevant for oblique shocks or for shocks
where the cosmic rays generate strong magnetic turbulence in the shock vicinity. The Kloeden-Platen predictor-corrector scheme proposed here
is such a scheme. We have demonstrated using simple simulations that the  Kloeden-Platen predictor-corrector scheme is significantly more accurate for small shock widths 
coupled with a strong gradient in the cosmic ray diffusivity.
For large shock widths (P\'eclet numbers larger than $\sim 0.25$), or when the gradient in the cosmic ray diffusivity is small, the two schemes produce comparable results
in terms of the accuracy of the momentum distribution that is obtained for the simulated particles.

Given the fact that the Kloeden-Platen predictor-corrector scheme is computationally about six times more expensive than the Cauchy-Euler scheme, one might consider
implementing a hybrid approach where one switches between the simpler Cauchy-Euler scheme and the Kloeden-Platen predictor-corrector scheme with the switch
based on the value of $|\Delta x_{\rm drift}/L_{\rm s}|$, the magnitude ratio of the drift term and the shock thickness
in the stochastic differential equation  (\ref{CES}) and the shock width. The results obtained here suggest that the KPPC scheme
is needed for sufficient accuracy whenever  $|\Delta x_{\rm drift}| \ga 4 \Delta x_{\rm adv}$ close to the shock. 

We have checked whether results with similar accuracy can be achieved with the Cauchy-Euler scheme, simply by reducing the time step until 
the computational expense is similar to that of the KPPC scheme with the larger time step. 
This turns out not to be the case. As an example we consider Model 2 for the case $\varepsilon = 0.04$. The analytical estimate for the slope is
$q_{\rm th} = 1.037$. We ran both schemes with $V_{1} \Delta t = 0.1$, $0.05$, $0.025$ and $0.0125$ in units where $L_{\rm s} = 1$, 
thus halving the time step each time. The KPPC scheme consistently returns (within errors due to Poisson noise)
a slope $q_{\rm KPPC} = 1.035$, quite close to the (approximate) theoretical result. 
Table 2 gives the corresponding result for the slope obtained with the CES, $q_{\rm CES}$. The slope still has a sizable error
even for the smallest time step, 8 times smaller than the largest step where the KPPC scheme already performs satisfactorily. 
The slope $q_{\rm CES}$, although decreasing as the time step gets smaller, remains consistently too large. 
We conclude that, for given computational expense, the KPPC scheme is still superior.

\begin{table}
\caption{Performance Cauchy-Euler scheme}
\label{table2}
\centering
\begin{tabular}{c  c c c c} \hline \hline
& & & & \\
Advective step $V_{1} \: \Delta t/L_{\rm s}$ & 0.1 & 0.05 & 0.025& $ 0.0125$\\
& & & & \\
Slope $q_{\rm CES}$ & 1.222 & 1.150 & 1.098 & 1.077 \\
& & & & \\ \hline \hline
\end{tabular}

\medskip
The slope $q$ of the simulated momentum distribution returned by the Cauchy-Euler scheme for Model 2 with $\varepsilon = 0.04$ for different time steps.\\
The theoretical slope equals $q_{\rm th} = 1.037$. 

\end{table}

We note in passing that the Kloeden-Platen predictor-corrector scheme may also be useful 
in other numerical applications of stochastic differential equations, such as the solution of the equations that describe stochastic particle
acceleration by waves (Fermi-II acceleration).  Here the equation for the evolution of the momentum distribution of the accelerating 
particles contains a momentum diffusion term that can be written in the form
(e.g. Melrose, 1980)
\be
	\left( \frac{\partial f}{\partial t} \right)_{\rm acc} = \frac{1}{p^2} \frac{\partial}{\partial p} \left(p^2 D_{p} \: \frac{\partial f}{\partial p} \right) \; , 
\ee
and a large drift term is unavoidable. The relevant drift velocity (in this case corresponding to the mean momentum gain) is
\be
	\left( \frac{{\rm d} p}{{\rm d}t} \right)_{\rm drift} = \frac{1}{p^{2}} \frac{\partial}{\partial p} \left(p^2 \: D_{\rm p} \right) \; .
\ee
In these expressions $D_{p}$ is the momentum diffusion coefficient.

\end{document}